\documentclass[
prd,aps,twocolumn,
superscriptaddress,showpacs
]{revtex4-2}

\usepackage{threeparttable}
\usepackage{aas_macros}
\usepackage{amsmath}
\usepackage{amssymb}
\usepackage{graphicx,float}
\usepackage[utf8]{inputenc}
\usepackage{booktabs}
\usepackage{array,multirow}
\usepackage{gensymb}
\usepackage{xcolor}
\usepackage{natbib}

\begin{document}

\title{\large An Evolving Leptonic Jet Model for Delayed Radio Flares
in Neutrino Blazars }
\author{Alina Kochocki}
\affiliation{\textit{Department of Physics and Astronomy, MSU, East Lansing, MI 48823, USA}}
\author{Xavier Rodrigues}
\affiliation{\textit{Université Paris Cité, CNRS, Astroparticule et Cosmologie, F-75013 Paris, France}}
\author{Nathan Whitehorn}
\affiliation{\textit{Department of Physics and Astronomy, MSU, East Lansing, MI 48823, USA}}

\begin{abstract}
The jets of blazar active galactic nuclei (AGN) are promising sites of hadron acceleration and subsequent neutrino production, owing to their extreme intrinsic power and high radiation density. Potential associations of IceCube neutrinos with blazars displaying delayed radio flares, such as TXS~0506+056 and PKS~1424+240, may support this scenario. However, the mechanisms and location of particle acceleration in the jet remain unclear. The 2017 IceCube event associated with TXS 0506+056 was concurrent with the initial peak of a flare in  gamma rays, followed by a longer flare at radio frequencies peaking three years later. State-of-the-art single-zone radiative source models focus on the compact high-energy-emitting region, and thus fail to describe this subsequent radio enhancement. In this work, we model the time-domain multi-frequency evolution of TXS 0506+056 by considering electron emission along the physically extended jet. We couple a numerical particle interaction framework with a dynamic description of the temporal and spatial jet evolution down to the parsec scale. We fit the model to multi-wavelength data, including multi-frequency radio light curves. The results suggest that the parsec-scale jet meets a population of older, cooled electrons near the edge of the observable radio core, establishing a causal relation between the initial high-energy emission and trailing radio activity.  This spatially and temporally resolved modeling approach can offer new insights into the physical conditions along the jet of flaring neutrino candidate blazars. 
\end{abstract}

\maketitle

\section{Introduction}
\label{sec:intro}

\subsection{High-energy astrophysical jets}

Blazars, active galactic nuclei (AGN) with relativistic jets pointing close to our line of sight, are among the brightest sources in the gamma-ray sky. Blazars emit a highly variable broadband electromagnetic spectrum from radio up to TeV gamma-rays. This gives evidence of efficient particle acceleration in the relativistic jet of AGN, making them natural laboratories of particle acceleration and transport in relativistic outflows~\citep{stat_plasma, plasma}.

The IceCube South Pole observatory has observed high-energy neutrinos in spatial association with some of these sources~\citep[e.g.]{txs_spectrum}. High-energy neutrinos are produced in inelastic proton-photon (p-gamma) and proton-proton (pp) interactions as sub-products of the hadronic decay of charged secondaries such as pions. Neutrino emission from blazars would therefore indicate that AGN jets accelerate protons or other cosmic-ray nuclei additionally to electrons. This multi-messenger approach is of special interest to address not only the open question of the chemical composition of AGN jets, but also the broader quest for the extragalactic cosmic-ray sources.

Hadronic interactions are typically efficient in compact regions with a high density of target photons or target matter. In the same compact region, we also expect efficient production of gamma-rays and X-rays, because a) hadronic processes also emit gamma rays through the electromagnetic decay of the neutral pion, as well as photons at other frequencies from secondary electron-positron pairs; and b) the same target photons also enable efficient inverse Compton scattering by primary electrons, providing an additional mechanism of gamma-ray emission. For these reasons, both gamma-ray and neutrino emission from blazar jets are expected to originate in the inner regions of blazar jets, typically within a few miliparsec of the supermassive black hole (SMBH), where the highest density of target photons can be found~\citep[e.g.]{Poutanen:2010he,Boettcher:2016xcw,Reimer:2018vvw,Rodrigues:2018tku}. 

Further from the central engine, the jet becomes less collimated as it often transitions from parabolical to conical~\citep{Kovalev:2019cue}. As the traveling plasma expands, the jet becomes more optically thin to synchrotron self-absorption (SSA), allowing lower-frequency radiation to escape. Radio observations of blazar jets can provide crucial insights into the jet's geometry, spatiotemporal dynamics, and chemical composition \citep{Clausen_Brown_2013, GABUZDA2003599, lister2007parsecscalejetenvironmentinteractionsagn, Makeev_2023}, in particular with the unmatched angular resolution of the very long-baseline interferometry (VLBI) technique. However, because the most compact zones often associated with high-energy emission are typically optically thick to SSA, radio observations often probe the larger parsec-scale structure \citep{Kovalev:2009bp,Boettcher:2013wxa,Lister:2016ojc}. Therefore, radio and high-energy messengers can probe distinct spatial scales and, likely, different underlying mechanisms of particle acceleration and transport~\cite{Plavin:2020mkf}. For this reason, connecting radio observations to the emission of high-energy radiation and neutrinos in a single theoretical framework is a challenging task~\citep[for a review, see][]{2019Galax...7...20B}.

\subsection{Flaring activity in neutrino blazar candidates}

In 2017, a likely muon neutrino with energy in excess of $\sim$300 TeV was observed from the direction of the blazar TXS 0506+056 \citep{Fermi_TXS_ATel}. The high energy of the event was above that expected for atmospheric background. This suggested that the event may have likely been an astrophysical neutrino, triggering a global multi-wavelength follow-up response \citep{txs_spectrum}. A gamma-ray flare also began in the months prior, increasing the source's luminosity by almost an order of magnitude before decaying over the following two years~\cite{txs_spectrum,ic_2018}. During this flare, the gamma-ray spectrum displayed a spectral index of $\sim$2 between 100 MeV to 100 GeV, before cutting off at the highest energies \citep{txs_spectrum}. Potential flaring activity was also observed in the X-ray band \citep{Acciari_2022}. 

A slow flare was also observed in the radio, consistent with synchrotron radiation emitted over a longer timescale, with onset during the gamma-ray flare and peaking three years after the IceCube event \citep{Chang_2022}. This radio activity was observed at multiple frequencies by the RATAN-600 telescope, showing an evolving spectral shape~\cite{Allakhverdyan_2023}. Similarly, a frequency-structured flare was observed at millimeter wavelengths by the Atacama Cosmology Telescope \citep{Abbasi_2026}. VLBI was used to study the varying morphology of the jet during this radio increase. The emission was found to originate from the radio core, consistent with a projected size of several parsecs \citep{Kun_2018}.  

The blazar PKS 1424+240, also associated with an excess of neutrino emission \citep{Aartsen_2020}, 
underwent a substantial gamma-ray flare between 100 MeV and 300 GeV from 2010 through 2014, followed by a significant radio flare from the core, starting in 2013 and peaking years later \citep{2023Symm...15..270K}. Radio data from the source has since been analyzed as indicating extreme Doppler beaming due to a down-the-barrel jet observation~\cite{Kovalev:2025kxf}. As there is no clear temporal association between neutrino emission and gamma-ray flaring for this source, we limit ourselves to TXS~0506+056 in this work.

Population studies have suggested that VLBI blazars spatially associated with IceCube neutrino events tend to have brighter parsec-scale radio cores than the general blazar population, and in some cases exhibit radio flaring activity around the time of the neutrino arrival \citep{Plavin:2020mkf, Kovalev:2023crn}. These findings have been interpreted as evidence that neutrino production occurs within a few parsecs of the central engine. Independent works using single-dish radio monitoring have also reported associations between neutrino arrival times and radio flares \citep{Hovatta:2020lor}, and additional statistical correlations between IceCube events and blazar catalogs selected by radio or gamma-ray brightness have been claimed by several groups \citep{Giommi:2020hbx,Buson:2022fyf}. Such  associations remain unconfirmed by IceCube stacking searches \citep{Abbasi:2024ewg,Bellenghi:2023yza,IceCube:2023htm,IceCubeCollaborationSS:2025jbi}, see also \citep{Franckowiak:2020qrq}, a tension that can partly be linked to the limited neutrino statistics. More recently, a characterization of the prominence of these gamma-radio delayed flares in a large sample of blazars has also been performed \citep{kochocki2026characterizinggammaradiodelayedflaring}. However, whether these delays arise from opacity effects near the jet base, or from the flare originating in a distinct region altogether, remains unresolved.

\begin{figure*}[ht]
\centering

\includegraphics[width=0.9\linewidth]{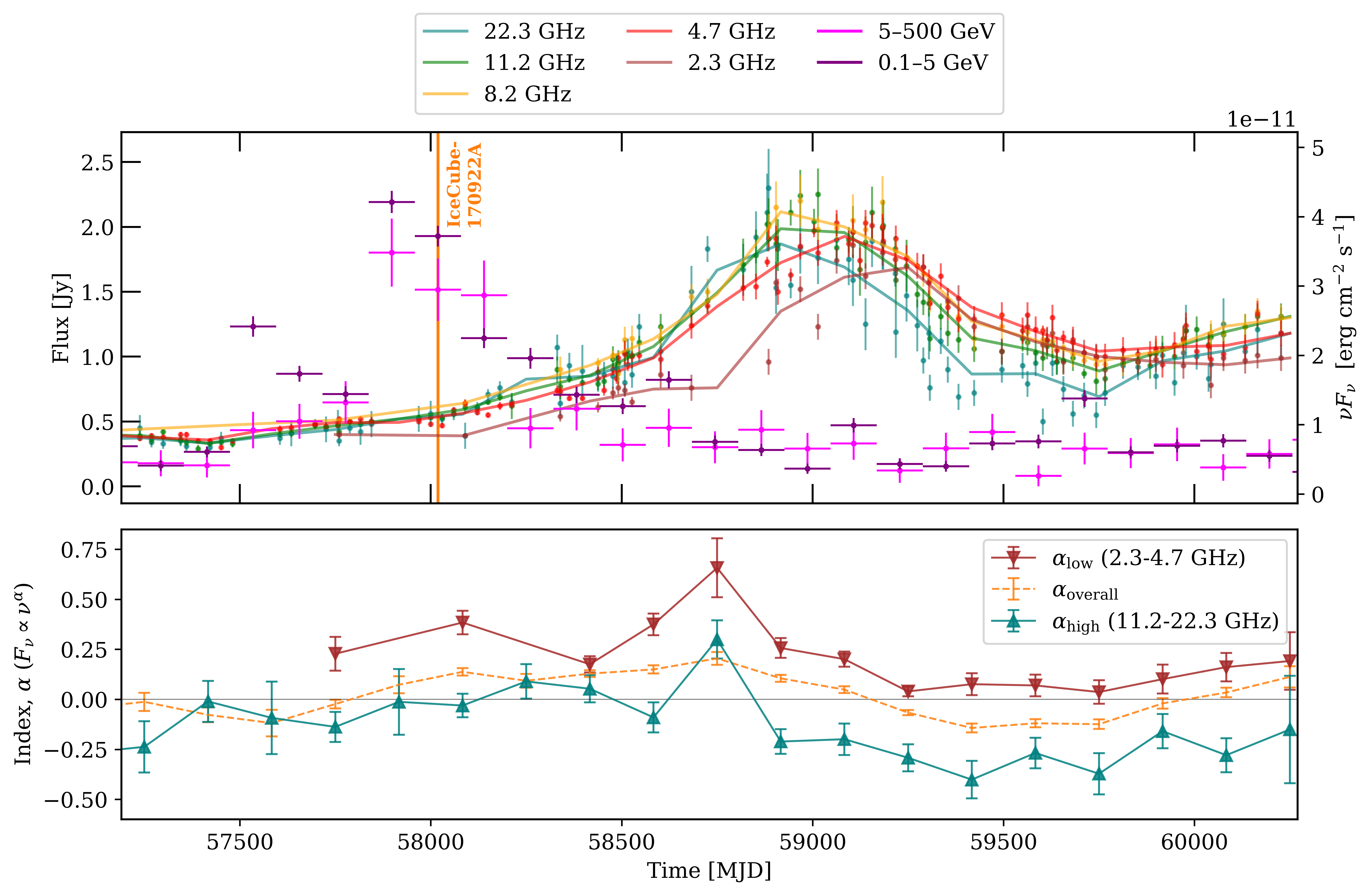 }

\caption{Multi-frequency light curves from TXS~0506+056. \textit{Upper plot:} in purple and magenta, we show the Fermi light curves analyzed in this work in two energy bands. The uncertainties represent symmetrized 68$\%$ confidence intervals. The orange vertical line shows the time of the 2017 IceCube-170822A alert event. In other colors we show time-domain RATAN-600 data at several frequencies \citep{Sotnikova_2022}. For visual reference, we plot for each frequency band a coarsely binned light curve (cf. main text). From visual inspection, we can already see some frequency-dependent variation in the flare shape, with lower frequencies generally peaking at later times. \textit{Lower plot:} best-fit spectral index of the radio data, using the same binning scheme, calculated between the two lowest frequencies (brown), the two highest frequencies (teal), and the overall index obtained by fitting a single power law across all frequencies (orange). At the lowest frequencies, the index becomes positive-valued during the flare's rising period, which our model explains as the sum of contributions from regions with different levels of self-absorption in the parsec-scale jet.}
\label{fig:txs}

\end{figure*}

\subsection{Delayed radio flares}

General radio flares may show a delay relative to gamma-ray emission for a variety of reasons. If gamma-ray and radio emission were linked to a single evolving blob of particles in the case of TXS 0506+056 and PKS 1420+240, a multi-year delay in the observer frame must be explained. An evolving self-absorption break from expansion may cause delayed radio flares on similar timescales \cite{Potter_2017}. Alternatively, as the blob expands, a general transition from Compton dominance to synchrotron dominance is expected with decreased density \citep{Boula_2018}. However, such an explanation depends on the geometry, kinematics and magnetic field evolution of the source. The widely used single-zone blazar modeling framework can generally describe well the steady-state broadband blazar emission in the optical, X-ray, and gamma-ray regimes~\citep[e.g.]{Boettcher:2013wxa,Rodrigues:2023vbv}, but fails to describe radio emission due to the compactness of the single zone. This severely limits the applicability of the single-zone model in interpreting multi-messenger correlations involving radio. Recent efforts to expand the framework from a single zone to a more sophisticated treatment of the physically extended jet are able to predict multi-wavelength light curves down to the radio frequencies ~\cite{Lucchini:2021scp,Zacharias:2022pea,Rodrigues:2025cpm}, which exceeds the   capabilities of the single-zone formalism. However, as we will argue, the fine structure of the time-domain radio data from TXS~0506+056 requires a more complex and dedicated modeling of the radio-emitting region.

In this work, we explore a possible causal relation between the 2017 gamma-ray flare and the delayed radio flare of TXS 0506+056. We consider that the jet accelerates a population of electrons at the miliparsec scale, emitting a short gamma-ray flare, and subsequently meets a configuration of electron clouds at the parsec scale, leading to fresh particle acceleration that describes the delayed radio flare. We apply the model to the 2017 flare of blazar TXS 0506+056, fitting the parameters to the time-domain multi-wavelength emission, including radio data. We show that the data are well described by propagation of the initially loaded jet material and subsequent interaction with a configuration of three clouds of cold electrons at a distance of several parsecs. By fitting the gamma-ray flare and the multi-frequency time-domain radio data, we constrain the particle density evolution along the jet and the spatial configuration of the parsec-scale clouds.

\section{Multi-wavelength Data}
\label{sec:data}

\subsection{Radio light curves from RATAN-600}
\label{sec:ratan}
We consider the temporal evolution of multi-frequency radio data from the RATAN-600 telescope, located at the Special Astrophysical Observatory (SAO) in Russia. The telescope utilizes a 576-meter diameter reflective ring with adjustable declination orientation. A secondary reflective system focuses to an array of receivers spanning 1-30 GHz in frequency. RATAN-600 has operated for several decades, with multiple campaigns focused on high-cadence observations of radio bright AGN. 

Radio light curves of TXS 0506+056 at frequencies 2.3, 4.7, 8.2, 11.2, and 22.3 GHz have been published by \citep{Sotnikova_2022}, shown as colored data points in the upper panel of Fig. \ref{fig:txs}. For visualization purposes, we overlay light curves for each band, shown in the corresponding color, obtained by interpolating from mean RATAN measurements over 30 large aggregate time bins, each spanning 167 days. Using this same coarse binning scheme, we show in the lower plot of Fig. \ref{fig:txs} the spectral index $\alpha$ (given by assuming a power-law $F_{\nu} = \nu^{\alpha}$ between two frequencies $\nu_1$ and $\nu_2$, cf. Eq. 1 of \citet{Sotnikova_2022}). The spectral index values in brown show the spectral index $\alpha_\mathrm{low}$ obtained between the two lowest-frequency bands, and in teal, $\alpha_\mathrm{high}$, between the two highest-frequency bands. In orange, we show the overall spectral index $\alpha_\mathrm{overall}$ given by fitting a power law to all bands at each time bin. As we do not expect a pure power law across all frequencies, the latter value merely serves as a rough indicator of the average value of the spectral index, which mostly falls between $\alpha_\mathrm{low}$ and $\alpha_\mathrm{high}$, as expected.

We can see from the lower plot of Fig. \ref{fig:txs} that there is a general hardening of the spectrum during the flare's rising period, followed by a steepening during the tail of the decreasing phase. As we discuss later on, in our model this spectral variability is described by the parsec-scale jet crossing partially overlapping regions with different electron densities, leading to a non-trivial time-dependent evolution of the total electron population. Throughout the entire event, the spectral index remains below the value expected for a SSA spectrum, $\alpha_{\textrm{SSA}} = 5/2$. As we will show, this evolution be described with a combination of regions with different optical thickness levels~\cite[see e.g. the discussion by][]{1980ApJ...238L.123C}. With such a combination of regions, the spectral index can vary in time and temporarily become positive while never reaching the value $\alpha_{\textrm{SSA}}$ corresponding to a completely optically thick emitting zone.

\subsection{Fermi-LAT gamma-ray light curves}

We present two light curves based on publicly available data from the Fermi-LAT gamma-ray telescope, in instrument with near-continuous cadence, sub-degree angular resolution between 1 GeV and 300 GeV, and degree-scale resolution between 20 MeV and 1 GeV. Owing to the high photon intensity, we perform a binned likelihood analysis, modeling TXS 0506+056 as a point source. We consider 45 equally spaced time intervals between December 11, 2009 and November 5, 2024, corresponding to 121-day bins. In each bin, we fit a power-law energy spectrum with free index and normalization to two energy ranges between 100 MeV and 5 GeV, and 5 GeV and 500 GeV. 

We have followed the standard analysis procedure outlined by Fermi: we downloaded the relevant photon data, spacecraft data and supplemental data products. We then filtered our data, selecting events of class `128' and type `3' which have a high probability of being either front or back converting photons. To remove photons from Earth's limb, we used a maximum zenith cut of 90 degrees. We only selected events with reconstructed energies between 100 MeV and 500 GeV for the analysis. 

We estimated the event counts map, livetime and exposure map as a function of space and energy. We assumed an instrument response function based on the first eight years of observation, Pass 8 P8R3 \citep{2012AAS...21914518A}. A model for galactic emission was used as well as an isotropic model for diffuse extragalactic emission. We used the Python package LATSourceModel to construct a list of contributing sources based on the LAT 14-year source catalog. If a source within 5 degrees of the target object exceeds a 5$\sigma$ significance threshold, these sources were fit with free spectral parameters. We constructed our binned likelihood analysis based on these expectations. 

We used a test statistic in analysis defined as the ratio between the null-result likelihood, $L_{0}$, and the best-fit likelihood, $L$, $TS = \textrm{2 ln}(L/L_{0})$. If a binned measurement had a test statistic less than 4, a 95$\%$ upper limit on the flux was placed. Otherwise, the symmetrized $68 \%$ confidence interval was reported. The results of this analysis are shown in magenta and purple in the upper panel of Fig. \ref{fig:txs}. 

Another nearby gamma-ray source in our search area, PKS 0502+049, was known to flare at energies below 1 GeV prior to 2016 \cite{Padovani_diss}, raising the possibility of source confusion. However, as our analysis focuses on the period after 2016, we do not expect this source to contribute substantially to our flux measurements. 

\begin{figure*}[ht]
    \centering
    \includegraphics[width=0.9\linewidth]{ 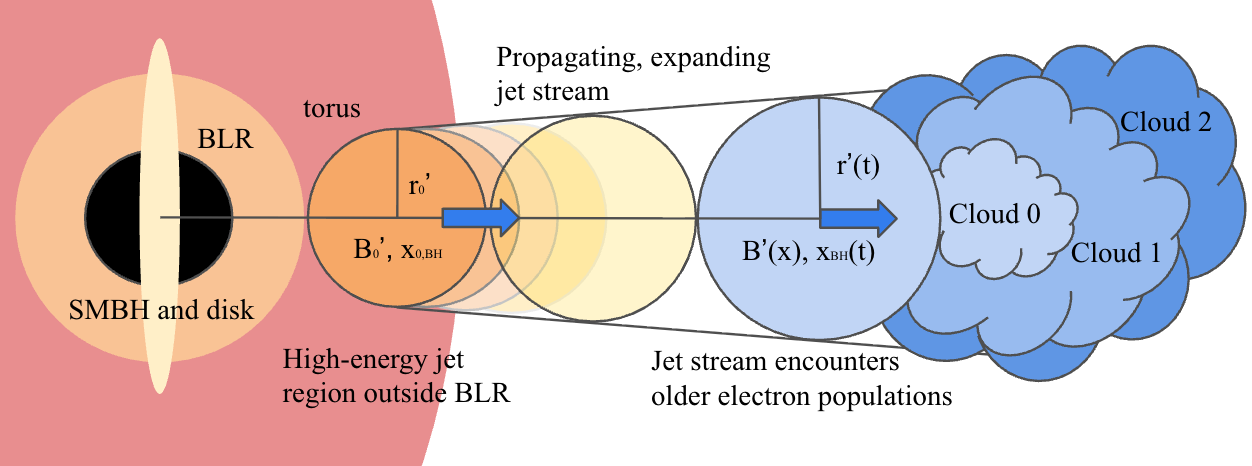 }
    \caption{A simplified diagram of our propagating, expanding and interacting jet model. Blobs of initial radius $r^\prime_{0}$ and distance to the SMBH $x_{0,\textrm{BH}}$ travel along the jet and expand at a constant rate, starting from just outside the BLR, where the gamma-ray flare originates. The magnetic field strength and the density of the background photon fields from the BLR and the dusty torus decrease with distance. At parsec scales, the leading blob interacts with a configuration of high-density clouds, accelerating fresh electrons whose synchrotron emission describes the observed multi-frequency radio flare.}
    \label{fig:jet}
\end{figure*}

\section{A Time-Dependent Model for Blazar Flares}

In this work, we wish to describe the gamma-ray and radio flares of TXS 0506+056 in a joint theoretical framework. To that end, we introduce a model that captures the interactions of a population of electrons in the compact inner jet and its subsequent propagation to the parsec-scale jet, where it accelerates fresh electrons from a cold population of electrons, modeled here as a sequence of clouds. We then fit the model, including the spatial profile of the electron clouds, by adding the predicted time-domain flare flux to a constant baseline steady-state flux and fitting the total to the time-domain gamma-ray and radio data in the period during and following the 2017 flare. In this section we describe the model assumptions, the methods involved in the numerical simulation, and the fitting procedure based on Markov chain Monte Carlo (MCMC) sampling.

\subsection{Time-dependent model setup}
\label{sec:simulation_setup}

Our time-domain model consists of a series of blobs launched from the inner jet, at a distance $x_{0,\textrm{BH}}$ from the SMBH, as represented schematically in orange in Fig. \ref{fig:jet}. At its original location $x_{0,\textrm{BH}}$, each blob has a radius $r^\prime_0$ and is permeated by a magnetic field of local strength $B^\prime_{0}$. \textit{Primed symbols refer to the local rest frame of the jet, symbols with a BH subscript refer to in the rest frame of the SMBH, and symbols without either marking refer to the observer's frame. Luminosities in the jet frame are also discussed without symbols. Thus, observed time intervals are given by $\Delta t = (1+z)  t^\prime/\delta_\mathrm{D}$, where $\delta_\mathrm{D}\,(\Gamma,\theta)$ is the Doppler factor of the jet, and $z$ is the source's redshift. Observed energies are given by $E = E^\prime \delta_\mathrm{D}/(1+z)$; and observed energy fluxes by $EF = E^\prime F^\prime  \delta^{4} / (1 + z)$.} Blazar TXS 0506+056 has been shown to host a broad line region \citep[BLR][]{Padovani_2019, Padovani_2022}, shown in Fig. \ref{fig:jet} in a lighter shade of orange, surrounding the SMBH. As listed in the upper section of Table \ref{tab:jet}, we adopt a value of $L_\mathrm{BLR}$ within the range derived by \citet{Padovani_2019}, and an according BLR radius $R_\mathrm{BLR}$ following \citet{Ghisellini_2009}. The dusty torus, emitting in the infrared range, is assumed to lie at a larger distance $r_\mathrm{torus}$, following the same reference (cf. Table \ref{tab:jet}).
We assume the blob to be initially located at a distance of a few times the BLR radius, $x_{0,\textrm{BH}}\gtrsim R_\mathrm{BLR}$, consistent with surveys of AGN jet formation and with radiative source models~\citep{Jorstad:2001eb,Kovalev:2009bp,Boettcher:2013wxa,Lister:2016ojc,Ghisellini_2009}.

In order to estimate the continuously varying time-domain emission, we simulate a series of blobs launched sequentially from $x_{0,\textrm{BH}}$ in temporal increments of $\Delta t_\mathrm{BH}=(2r_0^\prime/c)/\Gamma$. In each blob, we assume that pre-accelerated electrons are injected with a power-law energy spectrum. We model the interactions of these electrons and the resulting radiative cooling using the publicly available time-dependent numerical code AM$^{3}$ \citep{am3}. The code treats all particle species and magnetic fields as homogeneously and isotropically distributed in the plasma rest frame, and solves the time- and energy-dependent equations describing their cooling and respective electromagnetic emission, including non-linear cascades triggered by pair production. For a particle species $i$, the respective equation reads:
\begin{align}
\partial_{t} n^\prime_i(E^\prime, t^\prime) 
&= - \partial_{E^\prime} \left[ \dot{E}_i^\prime(E^\prime, t^\prime)n^\prime_i(E^\prime, t^\prime) \right] \nonumber \\
&\quad - \alpha_i^\prime(E^\prime, t^\prime)n^\prime_i(E^\prime, t^\prime) + Q_i^\prime(E^\prime,t^\prime),
\label{eq:pde}
\end{align}
where $n^\prime_i(E^\prime,t^\prime)$ is the time-dependent number density for particles of species $i$ and energy $E^\prime$, $\dot{E}^\prime(E^\prime, t^\prime)$ and $\alpha_i^\prime(E^\prime,t^\prime)$ represent continuous and discrete particle losses via interactions and escape, and $Q_i^\prime(E^\prime,t^\prime)$ represents particle source terms. As we neglect interactions of protons and other nuclei, our species are limited to electrons and photons, i.e. $i\in[\mathrm{e},\gamma]$, as well as positrons from pair-production processes, which behave similarly to electrons. The interactions include synchrotron emission, inverse Compton scattering (including off external photons from the BLR and the dusty torus), pair production from photon-photon annihilation, SSA, and adiabatic cooling of electrons and positrons owing to the jet expansion, as described further below. Physical escape from each blob is assumed to be energy-independent, representing a simple scenario of advective escape at light speed. 
To emulate the acceleration process, we inject electrons as a power law in energy $Q^\prime_\mathrm{e,inj} \propto E_\mathrm{e}^{^\prime2 - \gamma}, E^\prime_{\mathrm{min}} \leq E^\prime_\mathrm{e} < E^\prime_{\mathrm{max}}$ (cf. Eq. (\ref{eq:pde})), normalized in function of the total electron luminosity: $\int Q^\prime_\mathrm{e}dE^\prime=L_\mathrm{e}$. This normalization is fixed for each blob, and depends only on the blob's launch time $t$ in the observer's frame. Throughout this text, un-primed luminosities are assumed to denote values in the jet frame. We found empirically that the simplest temporal evolution function that can adequately describe the observed gamma-ray light curves is an asymmetric double-exponential function of time:
\begin{equation}
L_{\textrm{e}}(t) =
\begin{cases}
L_{0,\,\mathrm{jet}} \exp\!\left(\dfrac{t - T_{\textrm{peak}}}{ \lambda \, \sigma}\right), & t \le T_{\textrm{peak}}, \\[12pt]
L_{0,\,\mathrm{jet}} \exp\!\left(-\dfrac{t - T_{\textrm{peak}}}{ \sigma}\right), & t > T_{\textrm{peak}},
\end{cases}
\label{eq:elum}
\end{equation}
where $T_{\textrm{peak}}$ is the time of the flare peak in the observer's frame, and $\sigma$ and $\lambda$ represent the flare width and asymmetry scale, respectively. We inject electrons in each blob with a normalization following Eq. (\ref{eq:elum}) for a period of three times the light-crossing time of the blob, $\tau_{\textrm{steady}}^\prime = 3 \times r_{0}^\prime/c$. At that timescale, the electron population in the blob will have reached a steady state, owing to cooling effects and advective escape, as described above.
At that point, we stop the injection of primary electrons ($Q_\mathrm{e}^\prime=0$) and allow the blob to propagate downstream. As it propagates, the emission decreases exponentially as the remaining electrons cool and escape the expanding region.

The blob propagates downstream with a constant Lorentz factor $\Gamma$, expanding at speed $\eta c$. This expansion results in a reduction in the particle densities and the adiabatic cooling of the electrons and positrons, as mentioned previously (cf. \cite{Klinger_2024} for more details on the implementation of the blob expansion). In terms of the time $t^\prime$ in the rest frame of the blob, the size of the jet and distance to the SMBH are given by,
\begin{align}
    r^\prime(t) &= r^\prime_{0} + t^\prime \eta c, \\
    x_\mathrm{BH}(t) &= x_{0, \mathrm{BH}} + t_\mathrm{BH} \Gamma \beta c,
    \label{eq:r_and_x}
\end{align}
where $\beta c$ is the speed corresponding to the jet Lorentz factor $\Gamma$. As the blob moves away from the BLR, the densities of the photon fields from the BLR as seen in the blob frame decrease, following previous approaches~\citep{Ghisellini_2009,Rodrigues:2023vbv}.
We assume the magnetic field strength to evolve as a simple power-law in $x_\mathrm{BH}$:
\begin{equation}
    B^\prime(x) = B^\prime_{0} \times \left( \dfrac{x_\mathrm{BH}}{x_{0,\mathrm{BH}}} \right)^{-p}.
\end{equation}
For a magnetic field structure lying between purely poloidal and purely toroidal, the index $p$ lies in the range $1 \leq p \leq 2$ \citep{McKinney_2006}. As the results will show, our best-fit scenario favors a value closer to $p \gtrsim1$ \citep[cf. also Ref.][where a similar value of $p$ was also obtained in a proton synchrotron scenario]{Rodrigues:2025cpm}. At each time step, we update the magnetic field strength in the blob and the escape and expansion timescales according to the above relations, and we call AM$^3$ to evolve the distribution of electrons, positrons, and photons.

Since simulating the downstream propagation of all of these blobs throughout the entire duration of the flare would make the model fitting unnecessarily computation-intensive, we explicitly simulate the particle interactions in only a small number of representative blobs, and we assume that neighboring blobs emit identically to their nearest representative. For our $\sim$3.4 year injection period (as measured in the observer's frame), we found that explicitly simulating seven representative blobs, equally spaced time over this time period, provides sufficient temporal resolution to adequately sample the flare profile, given the rate of variation of the parameters along the jet.

To estimate the total time-dependent emission spectrum in the observer's frame, we sum the contributions of each of the blobs (the representative ones as well as the chain of contiguous copies), shifted by a time interval in the observer's frame of $\Delta t=(1+z)(2r_0^\prime/c)/\delta_\mathrm{D}$ compared to the previous blob in the chain, thus capturing their sequential launching. Finally, we account for the effect of photon interactions with the extragalactic background light (EBL), which leads to an additional attenuation of the observed gamma-ray flux. We adopt the frequency- and redshift-dependent attenuation factors as tabulated in the gammapy database \cite{Gammapy:2023gvb} assuming the EBL model by \cite{Franceschini:2008tp}. Following the above pipeline, we obtain the total multi-wavelength flux from the simulated inner jet flare as a function of photon frequency and time on the observer's frame, $dN_{\textrm{jet}}(E,t)/dEdt$.

The parameters describing the source environment and the jet structure described above are summarized in Table \ref{tab:jet}. In the upper section of the table, we show those parameters that we fix based on available information, following the respective reference. The remaining parameters are fitted to the time-domain data. Those parameters for which Table \ref{tab:jet} lists a single value were obtained with a by-eye fit to the mist robust features of the data; those for which a range is provided were  searched with our MCMC fit, described in Sec. \ref{sec:mcmc}.

\begin{table*}[t]
    \centering
    \begin{threeparttable}
    \caption{Model parameters describing the core AGN environment (upper section), the inner jet (middle section) and the parsec-scale clouds where fresh electrons are accelerated by the jet (lower section).}
    \label{tab:jet}
    \begin{tabular}{c|c|c}
    \hline
    \hline
    Parameter & Value & Description \\ \hline 
       $M_{\textrm{BH}}/M_{\odot}$ & $3 \times 10^{8}$ & BH mass \citep{Padovani_2019} \\
       $z$ & 0.3365 & Redshift \citep{txs_redshift} \\
       $\theta$/deg & 5 & Jet viewing angle \citep{Kun_2018}\\
        $L_{\textrm{disk}}$/(erg s$^{-1})$ & $8 \times 10^{44}$ & Accretion disk luminosity \citep{Padovani_2019}\\
       $L_{\textrm{BLR}}$/(erg s$^{-1}$) & $8 \times 10^{43}$ & BLR luminosity \\
       $r_{\textrm{BLR}}$/cm & $7 \times 10^{16}$ & BLR radius \\
       $r_{\textrm{torus}}$/cm & $2.5 \times 10^{18}$ & Torus radius \\
       $T_{\textrm{torus}}$/K & 500 & Torus temperature \\
       $\alpha_{\textrm{BLR}}$ & 0.1 & BLR covering \\
       $\alpha_{\textrm{torus}}$ & 0.3 & Torus covering \\ \hline
       $x_{0, \textrm{BH}}$/cm & $3.5 \times 10^{17}$ & Injection distance to the BH \\
       $\Gamma$ & $[1, 10]$ & Jet bulk Lorentz factor  \\
       $r_{0}$/($10^{15}\,\mathrm{cm}$)  & $[0.1, 5.0]$ &  Initial zone radius \\
       $\eta$ & $[0.05, 0.3]$ & Expansion rate in units of $c$ \\
       $B_{0}$ /G& $[0.1, 10]$ & Initial magnetic field strength \\
       $p$ & $[0.85, 2.1]$ & Magnetic field variation index\\
       $\log_{10}\left[ L_{0,\mathrm{jet}} / ( \mathrm{erg\,s^{-1}} ) \right]$ & $[41, 45]$ & Jet stream luminosity \\
       $\log_{10}\left[ E_{\textrm{min}}/(m_\mathrm{e} c^{2})\right]$ & $[1.0, 3.3]$ & Minimum injected electron energy\\
       $\log_{10}\left[ E_{\textrm{max}}/(m_\mathrm{e} c^{2})\right]$ & $[3.3, 8.7]$ & Maximum injected electron energy \\
       $\gamma$ & [1.7, 2.3] & Injected electron spectral index \\
       $T_{\textrm{peak}}$/yr & 1.35 & Jet flare peak, observer frame \\
       $\sigma$/yr & $[0.2,1.0]$ & Exponential flare scale, observer frame \\
       $\lambda$ & $[0.05,0.7]$ & Exponential flare asymmetry scale \\\hline
       $E_{\textrm{min, cloud}}/m_\mathrm{e} c^{2}$ &  $ 10$ & Minimum injected cloud electron energy\\
       $E_{\textrm{max, cloud}}/m_\mathrm{e} c^{2}$ &  $ 10^{3}$ & Maximum injected cloud electron energy \\
       $\log_{10}\left[ L_{0,\mathrm{cloud}\,0} / ( \mathrm{erg\,s^{-1}} ) \right]$ & $[42.0, 45.5]$ & Luminosity of cloud 0 \\
       $\log_{10}\left[ L_{0,\mathrm{cloud}\,1} / ( \mathrm{erg\,s^{-1}} ) \right]$ & $[42.0, 45.5]$ & Luminosity of cloud 1 \\
       $\log_{10}\left[ L_{0,\mathrm{cloud}\,2} / ( \mathrm{erg\,s^{-1}} ) \right]$ & $[42.0, 45.5]$ & Luminosity of cloud 2 \\
       $\gamma_\mathrm{cloud}$ & [1.0, 1.65] & Electron spectral index for clouds \\
       $T_{\mathrm{cloud}\,0}$/days & 1400 & Observer time of cloud center 0 \\
       $T_{\mathrm{cloud}\,1}$/days & 1525 & Observer time of cloud center 1 \\
       $T_{\mathrm{cloud}\,2}$/days & 1600 & Observer time of cloud center 2 \\
       $\epsilon$ & $[0.7, 1.3]$ & Cloud timing scale \\
       $\rho$/days & $[224, 416]$ & Cloud width scale \\
       $\xi_{0}$ & $[0.1, 0.35]$ & Relative size of cloud 0 \\
       $\xi_{1}$ & $[0.25, 1.0]$ & Relative size of cloud 1 \\
       $\omega_{l,0}$ & 1.11 & Left side asymmetry scale of cloud 0 \\
       $\omega_{l,1}$ & 0.77 & Left side asymmetry scale of cloud 1 \\
       $\omega_{l,2}$ & 0.2 & Left side asymmetry scale of cloud 2 \\ 
       $\omega_{r,0}$ & 0.2 & Right side asymmetry scale of cloud 0 \\
       $\omega_{r,1}$ & 0.3 & Right side asymmetry scale of cloud 1 \\
       $\omega_{r,2}$ & 0.2 & Right side asymmetry scale of cloud 2 \\ 
       \hline
       \hline
    \end{tabular}
    \begin{tablenotes}
    \small \item Note: for the parameters listed in the upper section of the table, we use values based on previous literature; for the parameters in the lower section, we fit their value to data together with the remaining model parameters, as described in the main text, searching within the range of values provided for each parameter. 
    \end{tablenotes}
    \end{threeparttable}
\end{table*}

We limit our model to leptonic processes, motivated by two observations: (1) in most single-zone leptohadronic models in the literature, where protons are typically accelerated up to sub-PeV energies, the contribution from hadronic processes to the multi-wavelength emission is negligible because of suppression of synchrotron emission relative to electrons by the large proton/electron mass ratio~\citep{Keivani:2018rnh,Gao:2018mnu,Cerruti:2018tmc,Acciari_2022,Petropoulou_2020}; (2) even if protons contribute significantly to the high-energy emission \citep[as predicted, for instance, by proton synchrotron models applied to this and other blazars][]{Cerruti:2014iwa,Rodrigues:2025cpm}, the radio emission coming from the parsec-scale jet is still likely to originate in synchrotron radiation by primary electrons, which is captured by a purely leptonic framework. Thus, although our model neglects direct emission from protons and other nuclei co-accelerated with electrons, the results can constrain the evolution of the jet dynamics and geometry, which are key elements for building a consistent picture of neutrino production in blazar flares.

\subsection{The parsec-scale jet and its environment}
\label{sec:model_radio_jet}

As the power in non-thermal electrons injected in the inner jet decreases exponentially, further particle injection in the parsec-scale jet is required to describe the radio flare. Reproducing the non-trivial behavior of the multi-frequency radio light curves, as introduced in Sec. \ref{sec:ratan}, requires a complex description of this particle injection profile. Our model explains the radio flare in the following way: when the first (leading) blob in the chain propagating from the inner jet reaches the parsec-scale jet, it runs into clouds of higher particle density (shown in shades of blue in Fig. \ref{fig:jet}), stationary in the radial direction in the SMBH frame. In the process, the jet accelerates a fraction of the electrons in these clouds to a non-thermal spectrum. We model this by injecting an additional electron power-law spectrum in the leading blob. Given that the acceleration mechanism and its efficiency in the parsec-scale jet is likely to differ from that in the inner jet, we allow for a different power-law index of the non-thermal electrons, $\gamma_\mathrm{cloud}$, compared to the inner jet. The injected electron population therefore follows $Q^\prime_\mathrm{e, cloud} \propto E_\mathrm{e}^{^\prime2 - \gamma_\mathrm{cloud}}, E^\prime_{\mathrm{min, cloud}} \leq E^\prime_\mathrm{e} < E^\prime_{\mathrm{max, cloud}}$. Because the minimum and maximum energies are strongly degenerate with $\gamma_\mathrm{cloud}$ and $L_\mathrm{e,cloud}$, we fix them for simplicity to $E_\mathrm{min, cloud}^\prime=10\,m_\mathrm{e}c^2$ and $E_\mathrm{max, cloud}^\prime=10^3\,m_\mathrm{e}c^2$. We disregard possible, additional, particle acceleration caused by the trailing blobs and not associated with the jet front. 

We found empirically that a fine-grained description of the multi-frequency radio data requires a minimum of three clouds of different sizes, partially overlapping in the radial direction. We parameterize the size of the clouds in the direction transverse to the jet motion as a fraction of the jet's local cross-sectional radius:
\begin{align}
r_\mathrm{cloud,0}&=\xi_0\,r,~\xi_0<\xi_1,\\ 
r_\mathrm{cloud,1}&=\xi_1\,r,~\xi_1<1,\\
r_\mathrm{cloud,2}&= r.
\label{eq:xi}
\end{align}
Clouds 0 and 1 can be smaller than the jet's cross-section, while cloud 2 is defined as the largest one, with a size of at least the jet's cross-section. Modeling for simplicity the cross-section of each cloud as circular, while crossing cloud 0(1) the jet sees a cross-section of $\pi [\xi_{0(1)}r(t)]^2$, while for cloud 2 it is simply $\pi r^2(t)$. Assuming the jet accelerates a constant fraction of the cloud particles, and given that the electron luminosity is dominated by the highest-energy electrons, the luminosity of non-thermal electrons at each point $x$ (in the SMBH rest frame)  follows approximately,
\begin{align}
L_{\mathrm{e,cloud}\,i}(x)&\propto\Gamma\beta c\,(\pi \, r_{\mathrm{cloud}\,i}^2)\, E_\mathrm{e}^\mathrm{max}\,n_\mathrm{e}(x),
\end{align}
where $n_\mathrm{e}$ is the cloud density at location $x$. As for simplicity we consider a constant value of $\Gamma$, the non-thermal electron luminosity injected in the leading blob at each point $x$ is simply proportional to the local cloud particle density. We model this profile as an asymmetric Gaussian for each cloud $i$:
\begin{equation}
L_{\textrm{e, cloud,}i}(x) = \begin{cases} L_{0,\textrm{cloud,}i} \exp\!\left[ \dfrac{1}{2}
\left(
\dfrac{x-\epsilon\,x_{\mathrm{cloud}\,i}}{ \omega_{l,i} \,\rho }
\right)^{2}
\right],
\\[6pt]
\qquad x \le x_{\mathrm{cloud}\,i}, \\[10pt]
L_{0,\textrm{cloud,}i} \exp\!\left[ -\dfrac{1}{2} \left( \dfrac{x-\epsilon\,x_{\mathrm{cloud}\,i}}{ \omega_{r,i} \,\rho } \right)^{2} \right], \\[6pt] \qquad x > x_{\mathrm{cloud}\,i}.
\end{cases}
\label{eq:electrons_cloud}
\end{equation}
Here, $x(t)$ is the position of the leading blob in the observer's frame and $\epsilon x_{\mathrm{cloud}\,i}$ is the central position of the $i$th cloud, where $\epsilon$ is a single nuisance parameter that shifts the spatial location of the entire cloud configuration along the jet. The leading blob crosses the center of a cloud located at $\epsilon x_{\mathrm{cloud}\,i}$ at a time $\epsilon T_{\mathrm{cloud}\,i}$; thus, these parameters directly influence the delay between the gamma-ray and radio flares.
The parameters $L_{0, \textrm{cloud,}i}$, $\rho$, and $\omega_{(l)r,i}$ correspond to the peak value, width, and asymmetry scale of the electron luminosity distribution in the leading blob caused by the crossing of each of the clouds.

To simulate the multi-wavelength emission from the leading blob as it crosses our three-cloud configuration, $dN_{\textrm{cloud}}(t)/dE_{\gamma}dt$, we simulate independently the particle interactions in the overlapping volume between the blob and each of the clouds, evolving them independently in time and summing their contributions to the total time-dependent emission. The implementation is described in further detail in Appendix~\ref{app:clouds}. 
We list the parameters describing the parsec-scale clouds and their respective range of tested values in the lower section of Table \ref{tab:jet}. 

\subsection{Monte Carlo model fitting}
\label{sec:mcmc}

Following the above jet modeling procedure, we obtain the total time-domain photon flux,
\begin{equation}
\begin{aligned}
\dfrac{dN_{\textrm{total}}(t)}{dE_{\gamma}dt}
&= \dfrac{ dN_{\textrm{jet}}(t)}{dE_{\gamma}dt}
 + \dfrac{ dN_{\textrm{cloud}}(t) }{dE_{\gamma}dt} \\
&\quad + \dfrac{dN_{\textrm{steady}}}{dE_{\gamma}dt},
\label{eq:fluxes}
\end{aligned}
\end{equation}
where we adopt the constant steady-state component, $dN_{\textrm{steady}}/{dE_{\gamma}dt}$, from a previous work describing the quiescent-state spectrum from the extended jet \cite{Rodrigues:2025cpm}, as described in further detail in Appendix \ref{app:steady}. We then fit the model by comparing this flux with the RATAN-600 and Fermi-LAT light curves for TXS 0506+056 introduced in Sec.~\ref{sec:data}.

We start with an initial parameter guess based on a by-eye fit of the broad data features, such as the approximate two-and-a-half year delay between the gamma and radio flares and the order-of-magnitude luminosity in both wavelengths. We then use this initial guess as a seed for MCMC sampling. We use 500 independent sampler chains, each performing 80 steps, evolved with the Metropolis-Hastings algorithm. We evaluate approximately $\sim$40,000 model realizations. We treat the first 30$\%$ of steps as a burn-in and parameter exploration.

We explore the parameter ranges listed in Tab. \ref{tab:jet}, informed by general principles: for example, we assume a bulk Lorentz factor $\Gamma \lesssim 10$, as determined by previous studies of radio morphology \citep{Kun_2018, Li_2020}, a spectral index $p \gtrsim 1$, as expected generally from Fermi processes, and a magnetic field strength in the inner jet of order of magnitude $B\sim$G, as determined in previous source models, as discussed previously. Because we restrict the sampling to the neighborhood of a single high-likelihood solution rather than the entire parameter space allowed by the priors, we can interpret the reported spread as a local likelihood-based uncertainty rather than a formal Bayesian credible interval derived from a global posterior sampling.

We assume measurement uncertainties are Gaussian and independent. The measurement, $y_{i,j}$, and uncertainty, $\sigma_{i,j}$, correspond to the $i$th photon energy, $E$, and $j$th observation time, $t$. Comparing the prediction, $f(\vec{\theta})$, of a sampled set of model parameters, $\vec{\theta}$, at a given time and photon energy, $t$ and $E$, we can evaluate the quality of the prediction through maximum likelihood estimation:
\begin{equation}
    \ln \mathcal{L}  = -\dfrac{1}{2} \sum_{i = 1}^{N_{E}} \sum_{j = 1}^{N_{t}(i)} \Bigg[ \dfrac{y_{ij} - f_{ij}(\vec{\theta}) }{\sigma_{ij} } \Bigg]^{2}.
\end{equation}
Finally, the nineteen parameters searched within their respective ranges are sampled with the corresponding uniform priors.

\begin{table}[]
    \centering
    \caption{Maximum-likelihood parameter values resulting from our MCMC sampling procedure. The range of values searched for each parameter is provided in in Tab. \ref{tab:jet}.}
    \label{tab:result}
    \begin{tabular}{c|c}
    \hline\hline
    Parameter & Highest-Likelihood Value  \\ \hline 
       $\Gamma$ & $4.54$   \\
       $r_{0}$/cm  & $0.35 \times 10^{15} $  \\
       $\eta$ & $0.14$  \\
       $B_{0}$/G & $1.15$  \\
       $p$ & $1.01$ \\ 
       $\log_{10}\left[ L_{0,\mathrm{jet}} / ( \mathrm{erg\,s^{-1}} ) \right]$ & $1.30 \times 10^{43}$  \\
       $E_{\textrm{min}}/m_\mathrm{e} c^{2}$ & $2.63\times10^2$ \\
       $E_{\textrm{max}}/m_\mathrm{e} c^{2}$ & $1.06 \times  10^{7}$ \\
       $\gamma$ & 1.94 \\
       $\sigma$/yr & $0.77$  \\
       $\lambda$ & $0.21$  \\
       $\log_{10}\left[ L_{0,\mathrm{cloud}\,0} / ( \mathrm{erg\,s^{-1}} ) \right]$ & $1.92 \times 10^{44}$  \\
       $\log_{10}\left[ L_{0,\mathrm{cloud}\,1} / ( \mathrm{erg\,s^{-1}} ) \right]$ & $2.41 \times 10^{44}$  \\
       $\log_{10}\left[ L_{0,\mathrm{cloud}\,2} / ( \mathrm{erg\,s^{-1}} ) \right]$ &  $2.00 \times 10^{44}$  \\
       $\gamma_\mathrm{cloud}$ & 1.11  \\
       $\epsilon$ & $1.01$  \\
       $\rho$/days  & 281  \\
       $\xi_{0}$ & $0.21$  \\
       $\xi_{1}$ & $0.34$  \\
       \hline\hline
    \end{tabular}
\end{table}

\begin{figure*}[ht]
    \centering
    \includegraphics[width=\linewidth]{ 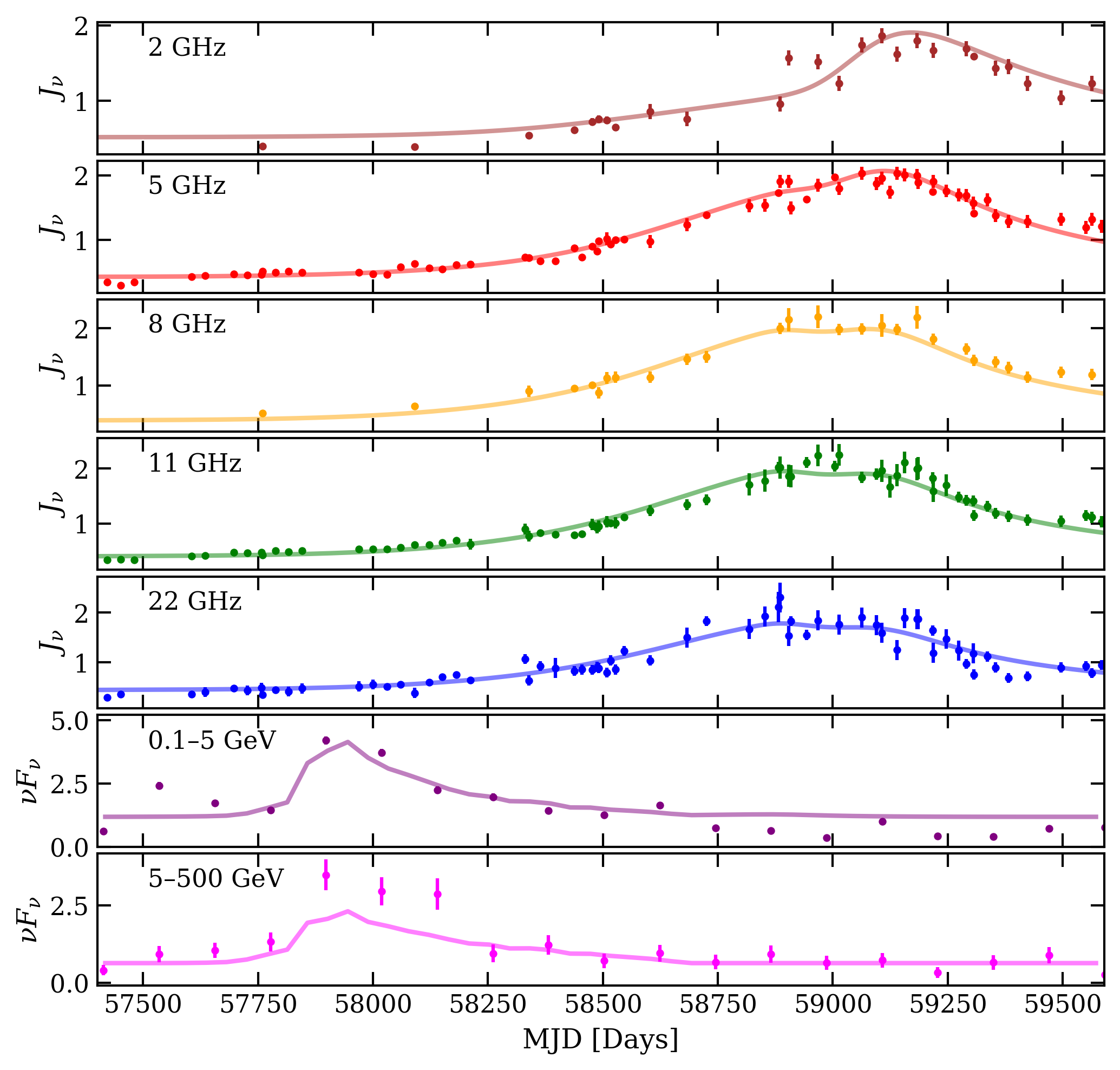 }
    \caption{Light curves of the highest-likelihood jet model explored in Monte Carlo sampling. We provide a comparison of our Fermi-LAT light curves in two energy bands and public RATAN-600 data across several frequencies with our highest-likelihood model result. We find good agreement for both the initial gamma-ray flare and later radio activity. Importantly, the frequency-dependent evolution of the radio flare is well described with our multiple zone parameterization.}
    \label{fig:res_curves}
\end{figure*}

\begin{figure}[ht]
    \centering
    \includegraphics[width=\linewidth]{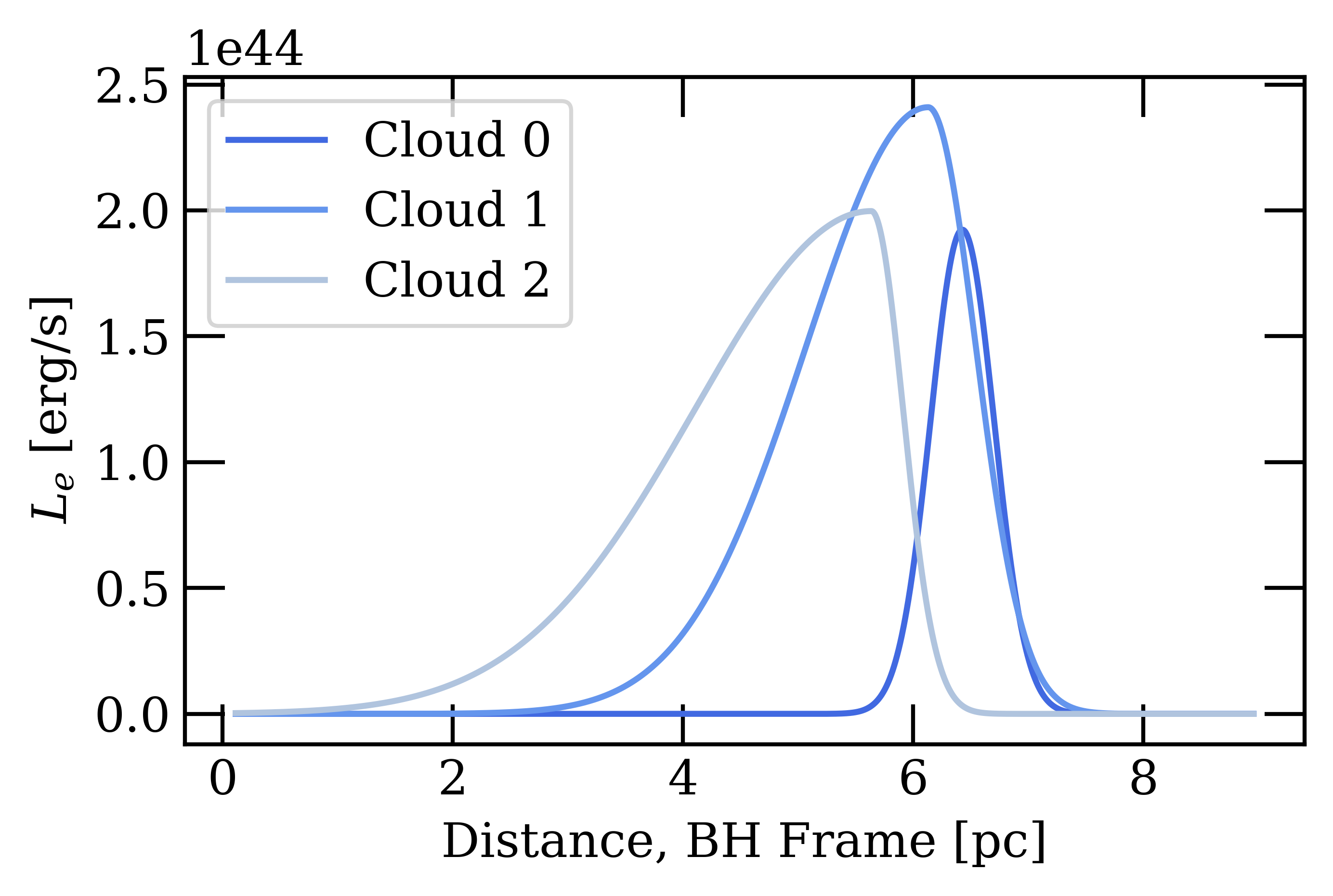}
    \caption{Best-fit luminosity profiles of the electron clouds driving the radio flare, as a function of radial distance $x_\mathrm{BH}$. These profiles reflect the free parameters $L_{0, \textrm{cloud}\,0}, L_{0, \textrm{cloud}\,1}, L_{0, \textrm{cloud}\,2}$, as well as $\epsilon, \rho, \xi_{0}$ and $\xi_{1}$ (cf. Table~\ref{tab:jet}).}
    \label{fig:cloud_fit}
\end{figure}

\section{Results}
\label{sec:results}

We show the best-fit multi-wavelength light curves in the observer's frame in Fig. \ref{fig:res_curves}: the upper five panels show the multi-frequency radio light curves, while the two lower panels show the gamma-ray light curve in the two analyzed frequency bands. The best-fit values of the parameters optimized through the MCMC procedure are listed in Table \ref{tab:result}.

As we can see in Fig. \ref{fig:res_curves}, the model captures the non-trivial frequency dependence of the radio flare, owing to the complex spatial profile of the clouds swept by the parsec-scale jet (cf. Sec. \ref{sec:discussion} for a qualitative comparison with previous works). At 2~GHz, shown in the upper panel, the model correctly describes a late peak, about 3.5 years after the peak of the gamma-ray flare. In contrast, higher radio frequencies display an earlier rise phase with a more temporally extended flare containing two sub-peaks; as we discuss below, this is due to the jet crossing the earlier clouds. The decay phase is shorter than the rise phase, with a duration of only about a year, and is roughly simultaneous across all frequencies, in agreement with the data.

To physically interpret the predicted radio light curves, we show in Fig. \ref{fig:cloud_fit} the best-fit spatial distribution of the electron injection luminosity in the parsec-scale jet resulting from the crossing of each of the clouds. As we discussed in Eq. (\ref{eq:electrons_cloud}), this quantity is proportional to each cloud's spatial density profile, as the jet is assumed to have constant speed and accelerate a constant fraction of the particles that it picks up. We can see that the clouds lie at a radial distance of between 3 and 7~pc. All clouds impinge on the jet a peak non-thermal electron power of $\sim2\times10^{44}\,\mathrm{erg/s}$ (cf. also Tab. \ref{tab:result}). The most upstream cloud is the most extended one in the radial direction, starting at about 2~pc and peaking just below 6~pc. At the same time, it has a relatively small transversal radius (cf. $\xi_0$ Tab.~\ref{tab:result}), making the emission region relatively compact. As our leading blob crosses this first cloud, the synchrotron emission from the freshly accelerated electrons describes the slow early rise observed between 5 and 22 GHz. This is controlled by the parameter $\omega_{l,0}$ (Tab.~\ref{tab:jet}). Owing to the compactness of the emitting region, the escaping photons are partially self-absorbed; this explains the fact that the 2~GHz flux does not initially increase as fast as at higher frequencies.

The jet blob then crosses the second and third clouds, both peaking just beyond 6~pc. While the second cloud is also compact (cf. $\xi_1$ Tab. \ref{tab:result}), the third cloud is larger, as described in Sec. \ref{sec:model_radio_jet}, and the corresponding emitting region spans the entire jet cross section. This makes the region more optically thin, leading to the late 2~GHz flare. The simultaneous emission at higher frequencies contains a contribution from both cloud 2 and cloud 1. 

Finally, as the blob moves away from the cloud complex, the radio flux decreases simultaneously at all frequencies. This decay time is determined by the parameters $\omega_{r,1}$ and $\omega_{r,2}$, which describe the decrease of the spatial density profile of the last two clouds (cf. Tab. \ref{tab:jet}). Altogether, the model predicts that the bulk of the radio emission during the flare is entirely confined to the innermost $\sim$7~pc, roughly compatible with measurements of the VLBI core~\cite{Kun:2018zin,Ros:2019bgo}.  

\section{Discussion}

\subsection{Advancements compared to previous works}
\label{sec:discussion}
As we discussed in Sec. \ref{sec:intro}, the widely-used single-zone blazar model cannot simultaneously describe the high-energy and radio emission. This is due to the fact that high-energy emission requires a high density of target radiation, implying a high optical thickness to radio wavelengths owing to the SSA process. In light of this, it is evident that a more complex model is necessary to connect these two ends of the electromagnetic (and multi-messenger) spectrum. With the present model, we expand on the one-zone model by considering a) a continuous stream of blobs in the inner jet, to predict a time-domain gamma-ray light curve rather than simply a steady-state spectrum; b) fresh particle acceleration in the parsec-scale jet additionally to the inner jet, to explain radio emission down to 2~GHz; and c) a profile of parsec-scale clouds, to describe the detailed time-domain structure of the multi-frequency radio light curves. The cost of this higher complexity is a higher number of parameters: as we can see from the two lower sections of Tab. \ref{tab:jet}, our model has a total of 30~parameters compared to the $\sim$10 of a typical one-zone model. In that sense, the present model serves a more descriptive role, as it uses more detailed time-domain data to constrain the parsec-scale environment.

Another recent work has described the emission from TXS~0506+056 with an extended jet model that treats particle acceleration and diffusion as a continuous process driven by stochastic scattering off a turbulent magnetic field along the jet~\citep{Rodrigues:2025cpm}. Unlike the current work, that model has a comparable number of parameters to a typical single-zone model, owing to the continuous treatment of the jet from the BLR out to the multi-parsec scales and the simplified treatment of its geometry. Although that model describes radio emission down to lower frequencies compared to single-zone models (see references provided in Sec. \ref{sec:intro}), it still under-describes the total radio flux below a few tens of GHz. That suggests that treating the extended jet as a continuous, homogeneous flow with continuous particle acceleration fails to capture particular features necessary to describe the radio emission. Such features can be a) an increase in the number of non-thermal particles, as considered in the present work, b) an additional acceleration mechanism that becomes more efficient in the low-magnetization parsec-scale environment, such as shock-driven acceleration, or c) a more optically thin region that allows for the escape of low-frequency photons. The latter can be ascribed, among other possibilities, to a less collimated parsec-scale jet, as observed in some radio galaxies, or an extended structure in the poloidal direction, such as a slower sheath surrounding the faster jet, as suggested by another recent work \cite[][cf. discussion in Sec. \ref{sec:limitations}]{Kovalev:2026fba}. As we discussed previously in the context of the present model, any of these extensions necessarily increase the model's complexity.

Another recent paper, published during the final stages of preparation of the present work, has also modeled the delayed radio flare of TXS~0506+056 \citep{stathopoulos2026delayedradioflaresneutrinoassociated}. That model also explains the delayed radio flare of TXS 0506+056 as originating in additional particle acceleration in the parsec-scale jet, but with a few significant differences to the present one. First, the timing of the radio flare and its decay phase is attributed to the deceleration of the jet, while our model ascribed this to the spatial profile of the parsec-scale clouds. Secondly, the model by \citep{stathopoulos2026delayedradioflaresneutrinoassociated} predicts a simple exponential profile of the radio flare, which peaks simultaneously  across all radio frequencies. In contrast, the present model describes the non-trivial frequency-dependent behavior of the radio light curve, including the frequency dependence of the peak time. As discussed in the previous section, this is possible thanks to the spatial superposition of multiple electron populations, and cannot easily be reproduced with simpler frameworks.

\subsection{Implications for neutrino astrophysics}

By connecting the 2017 gamma-ray flare of TXS~0506+056 with the subsequent multi-frequency radio flare, our model suggests a causal relationship between the two: the perturbation in the inner jet that leads to the high-energy emission propagates out to the parsec scales, where it accelerates fresh particles lying in the environment, causing the observed delayed radio flare. In the broader landscape of neutrino candidate blazars, delayed radio flares have been noted for TXS~0506+056 as well as PKS~1424+240, both intermediate-peaked synchrotron blazars \cite{kochocki2026characterizinggammaradiodelayedflaring}. As we discussed in Sec. \ref{sec:intro}, numerous recent works have suggested associations between high-energy IceCube events and jetted AGN, although IceCube stacking searches, statistically weighting sources either by their gamma-ray or radio flux, have only placed constraints on the neutrino flux from the population~\citep{Abbasi:2024ewg,Bellenghi:2023yza,IceCube:2023htm}. This implies that while individual high-energy IceCube events may in fact originate in blazar jets, blazars cannot all be equally strong neutrino emitters~\cite{IceCubeCollaborationSS:2025jbi}.

One possibility is that different sources may vary considerably in terms of the composition of the inner jet or the locations of efficient particle acceleration, leading to a non-trivial relationship between the multi-wavelength emission and the accompanying neutrino emission. While current models cannot yet produce such detailed constraints, with our current approach we show that available time-domain data can constrain the source environment, which is a crucial element in the ultimate multi-messenger picture of jetted AGN.     

As we argue below, future studies will have to include more flexible jet geometries and other source properties. Explicitly including hadronic processes in such environments will then also allow a more direct  test of neutrino production in different source candidates. This combination of simulations and time-domain multi-messenger data will hopefully shed light into a variety of source properties that cannot be captured with current frameworks. 

\subsection{Model limitations}
\label{sec:limitations}

Our model requires the evaluation of the multi-wavelength emission from an evolving particle population over several years in the observer's frame, which is resource-intensive. This limitation constrains both the size of the parameter space and the viable duration of the MCMC sampling. To reduce these computational resource requirements as much as possible, requirements, we used an initial guess as a seed for the search based on a by-eye fit of the most robust features; we then performed MCMC sampling in a small parameter space around that model selection to demonstrate the local likelihood structure. While the best-fit result describes well the data, we cannot easily exclude other solutions owing to this limitation. We encourage other works to study further these degeneracies and potentially find new solutions that describe the data at an equal level of detail in the time domain.

Another limitation is the parameter degeneracy intrinsic to any leptohadronic model. For example, there is a degree of degeneracy between the bulk Lorentz factor $\Gamma$, the observation angle $\theta$, and the distance between the locations of the gamma-ray and radio flares. These geometric uncertainties are compounded by known degeneracies intrinsic to the leptonic and leptohadronic frameworks themselves \citep[see e.g. Fig. 3 of][]{Rodrigues:2026kpf}, for example between the magnetic field strength $B_{0}$ and the maximum energy $E_\mathrm{max}$ and the luminosity $L_{0,\mathrm{jet}}$ of electrons accelerated in the inner jet. 

While our model only accounts for the emission from non-thermal electrons, it is likely that the jet also accelerates a fraction of protons. In most single-zone models of TXS~0506+056 (see Sec. \ref{sec:intro} and references therein), the bulk of the gamma-ray flux is ascribed to electron emission. This motivates our choice of neglecting the emission from non-thermal protons, as that would introduce additional free parameters describing the proton population. At the same time, in some leptohadronic models, protons can contribute partly to the gamma-ray flux \cite{Rodrigues:2024fhu}, or even dominate it completely \cite{Cerruti:2018tmc,Rodrigues:2025cpm}, and further research is necessary to verify the current findings in such models. A hybrid model of neutrino blazar candidates suggests that electron emission may dominate the gamma-ray flux below some 100~GeV, and proton emission above ~\cite[e.g.]{Cerruti:2018tmc,Rodrigues:2024fhu}. Such a scenario is consistent with the fact that the present model under-predicts the total gamma-ray flux observed during the flare in the 5-500 GeV band (lower panel of Fig. \ref{fig:res_curves}). This also requires further investigation by including proton acceleration in the model proposed here. As it stands, the model presents a basic description of electron emission, and our results can constrain the dynamics of the jet with a minimal parameter space dimensionality. Future work incorporating the role of non-thermal protons will allow for self-consistent estimates of neutrino production during the 2017 flare under different acceleration scenarios. 

It is also likely that more complex structures are present in the jet. A recent work on VLBI observations of TXS~0506+056 has suggested the presence of a slow sheath surrounding an ultra-fast spine, ~\cite{Kovalev:2026fba}. This implies that the bulk of the emission associated with the spatially evolving radio features should originate at a distance to the black hole comparable to that of the gamma-ray-emitting region, simply further away from the jet axis. In comparison, in the present work where the jet is modeled as a one-dimensional flow, the radio emission is predicted to originate at considerably larger distances, between 2 and 7~pc. While at face value this is at odds with the above observations, the intrinsic degeneracies mentioned above make the absolute value of the position of the radio emission difficult to constrain definitively in this framework. On the other hand, the relative configuration and the size of the radio emission zones are more robust features of the model, as they are constrained by the multi-frequency behavior of the radio light curve and the implied temporal dependence of the optical thickness of the multiple emitting regions. 

Even within a one-dimensional framework, variations in the orientation of the jet~\cite[cf. Fig. 2 of Ref.][]{Ros:2019bgo} can lead to varying degrees of Doppler boost along the spatial coordinate, introducing an additional complexity compared to the fixed viewing angle we assumed here. Some authors have also suggested indications of jet curvature or precession in TXS~0506+056 and other neutrino blazar candidates \citep{2019A&A...630A.103B, Kun_2018}, or the effect of gravitational lenses interposed between the source and the observer, potentially amplifying the intrinsic emission \cite{Britzen:2025bww,Britzen:2026:lensing}. In light of these and previous considerations, it is clear that the present model is still limited in the treatment of the source physics and geometry, in spite of its additional complexities beyond the single-zone framework. Integrating further data, particularly information provided by VLBI measurements, may yield important additional constraints that cannot be captured by the model in its current form. We leave these model extensions for future work.

\subsection{Conclusion}

We presented a time-domain model of the 2017 gamma-ray flare and the subsequent multi-frequency radio flare of blazar TXS 0506+056. We considered a scenario in which a propagating and expanding inner jet, responsible for the gamma-ray flare, interacts with a complex of clouds as it reaches the parsec scale, accelerating fresh electrons and explaining the delayed radio flare. We fit the model to publicly available data gamma-ray light curves from Fermi-LAT and multi-frequency radio light curves from RATAN-600, using a local likelihood minimization driven by MCMC sampling. Fitting to the time-domain data has allowed us to constrain key model parameters such as the location and spatial distribution of the electron clouds responsible for the radio flare. As non-thermal electrons are typically expected to dominate the radio emission, and even the gamma-ray emission in most leptohadronic models, we have neglected contributions from non-relativistic protons.

With a best-fit Doppler factor of $\delta_\mathrm{D}=7.8$ for a viewing angle of 5$^\circ$ (both assumed constant), the cloud distribution responsible for the radio emission is constrained to lie between 2 and 7~pc. The first cloud, lying between 2 and 5~pc, has the smallest cross-sectional overlap with the jet, yielding a compact emission zone. This explains the early rise in the radio flux at and above 5~GHz, while at 2~GHz the emission is absorbed via SSA. As the jet crosses the second and third clouds, the emission at and above 5~GHz undergoes a double peak in the time domain, about 2.6 and 3.2~years after the gamma-ray flare. As the third cloud intersects with the entire jet cross-section, the emission zone is the most optically thin. This describes the long delay observed in the 2~GHz band, where the peak occurs only about 3.4~years after the gamma-ray flare.

While the model does not directly capture neutrino emission, it constrains the conditions necessary to produce a non-trivial multi-frequency radio flare. We thus probe source properties that may be key to unlocking a broader and more consistent picture of multi-messenger emission from blazars and jetted AGN at large. More broadly, our results show that fitting time-domain multi-wavelength data allows us to increase the complexity of source models, pushing them beyond the state-of-the-art single-zone framework. Including more detailed observational data, such as time-domain VLBI imaging, may eventually allow the inclusion of even more complex geometries, enabling new probes of the physics of AGN jets.

\begin{acknowledgments}
A.K. thanks support from NSF grant PHY-2237581. X.R. acknowledges support by the investment program ‘France 2030’ launched by the French Government and implemented by the University Paris Cité as part of its program ‘Initiative d’excellence’ IdEx (ANR-18-IDEX-0001), which also funded the HERMES: multi-messengers of the Earth and the Universe project that contributed to this work. This paper makes use of publicly available \textit{Fermi}-LAT data provided online by the \url{https://fermi.gsfc.nasa.gov/ssc/data/access/} Fermi Science Support Center. This work was supported in part through computational resources and services provided by the Institute for Cyber-Enabled Research at Michigan State University.
\end{acknowledgments}

\appendix

\section{Numerical approach to the jet-cloud simulation}
\label{app:clouds}

As we explained in the main text, our model describes the radio flare as the result of the interaction of the parsec-scale jet with clumps of higher particle density, which we refer to in a generic manner as ``clouds'', leading to a temporary enhancement in the luminosity of the accelerated particles. For the sake of simplicity, we consider only particle emission from the leading blob, i.e. the first of the blobs emitted from the inner jet, as it reaches the parsec-scale jet. 

We found empirically that reproducing the non-trivial features of the multi-frequency radio light curves requires a minimum of three clouds. We consider three clouds, modeled as three electron distributions along the radial direction (i.e. the jet's direction of motion), given by Eq.~(\ref{eq:electrons_cloud}). Electron emission is assessed in the overlapping volume between the passing jet blob and each of the clouds. In the poloidal plane (transverse to the jet direction), the overlapping volume between the jet blob and the clouds 0, 1, and 2 has a size given respectively by  $\xi_{0}r$, $\xi_{1}r$, and $r$ [Eq. (\ref{eq:xi})]. In our best-fit scenario, we have $\xi_0<\xi_1$ (Table \ref{tab:result}), so cloud 0 is the smallest, cloud 1 is larger, and cloud 2 is the largest by definition, as it spans the entire jet cross-section.  

For simplicity, and because our jet model is coaxially symmetric along the radial direction, we assume the clouds to also be coaxial along this direction. Thus, in the sections where clouds overlap, the smaller clouds are embedded inside the larger ones. In this configuration, there are only three qualitatively different volumes swept by the passing blob: a volume containing particles from only the largest cloud ($V_a$), a volume of overlap between the two largest clouds ($V_b$), and a volume of overlap between all three clouds ($V_c$). As the blob itself is spherical, we can approximate each of these volumes as also spherical, yielding: 
\begin{align}
    V_{a}(t) &= \dfrac{3}{4} \pi \, r(t)^{3} (1 - \xi_{1} )^{3}, \\
    V_{b}(t) &= \dfrac{3}{4} \pi \, r(t)^{3} (\xi_{1} - \xi_{0})^{3}, \\
    V_{c}(t) &= \dfrac{3}{4} \pi \, r(t)^{3} \xi_{0} ^{3}.
    \label{eq:cloud_volumes}
\end{align}
In our AM$^3$-based numerical approach, we treat this system by modeling the time evolution of these three regions independently. Since we assume the particles to be distributed uniformly throughout each cloud, the total electron luminosity in each independent zone is given simply by the sum of the contributions from each cloud:
\begin{align}
L_{a}(t) &= L_{\mathrm{e,cloud}\,0}(t)\,(1-\xi_{1}^{3}), \\[6pt]
L_{b}(t) &= L_{\mathrm{e,cloud}\,0}(t)\,(\xi_{1}^{3}-\xi_{0}^{3}) \nonumber \\
&\quad + L_{\mathrm{e,cloud}\,1}(t)\,(\xi_{1}^{3}-\xi_{0}^{3})/\xi_{1}^{3}, \\[6pt]
L_{c}(t) &= L_{\mathrm{e,cloud}\,0}(t)\,\xi_{0}^{3} 
+ L_{\mathrm{e,cloud}\,1}(t)\,\xi_{0}^{3}/\xi_{1}^{3} \nonumber \\
&\quad + L_{\mathrm{e,cloud}\,2}(t).
\end{align}
The resulting time-dependent flux emitted as a result of the blob crossing the entire cloud complex is then estimated as
\begin{equation}
    \dfrac{dN_{\textrm{cloud}}(t)}{dE_{\gamma}dt} = \dfrac{dN_{\textrm{a}}(t)}{dE_{\gamma}dt} + \dfrac{dN_{\textrm{b}}(t)}{dE_{\gamma}dt} + \dfrac{dN_{\textrm{c}}(t)}{dE_{\gamma}dt}.
\end{equation}
We note that the actual overlap between the jet and the different clouds it crosses may deviate from this simple, coaxially symmetric picture. Such deviations in the true geometry of the system can in principle be captured by a more elaborate model. However, for the purpose of the present work, a correct characterization of the evolving compactness of the emitting region is  more relevant than an exact geometric description.

\section{Parameter posterior distributions}

\begin{figure*}[ht]
    \centering
    \includegraphics[width=\linewidth]{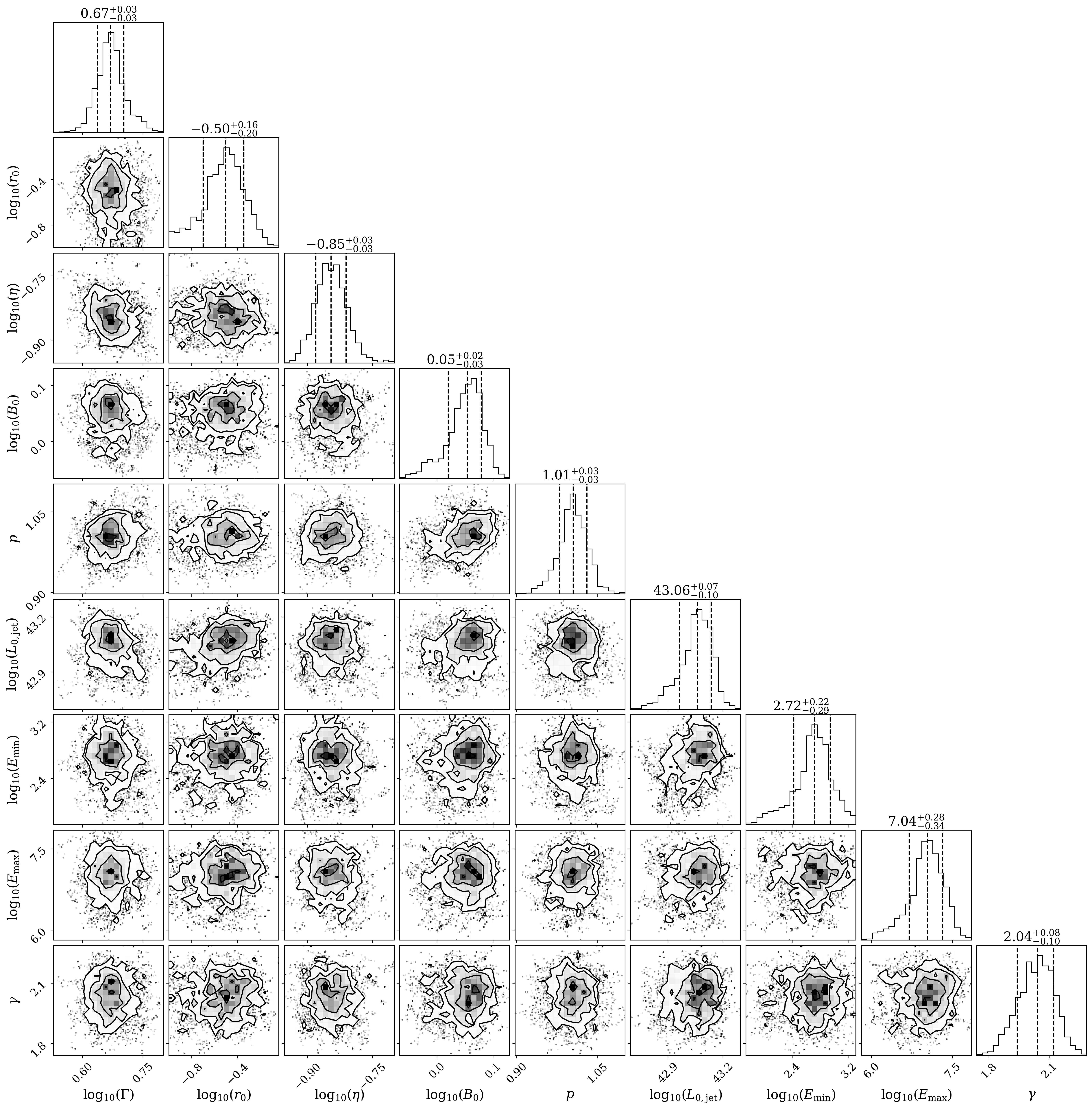  }
    \caption{Parameter values relevant to the jet stream explored through MCMC sampling. As our model has a high computational cost, we evaluate approximately $\sim$40,000 samples. The result is intended to represent exploration of the parameter space near a local, high-likelihood solution, as opposed to fully converged or globally representative Bayesian credible intervals. Contours correspond to the 1, 2, and 3$\sigma$ credible regions in two dimensions. }
    \label{fig:res_triangle}
\end{figure*}

For the eight parameters describing the inner jet (cf. Sec. \ref{sec:simulation_setup}), we show their sampled posterior distributions in Fig. \ref{fig:res_triangle}. 

\begin{figure*}[ht]
    \centering
    \includegraphics[width=\linewidth]{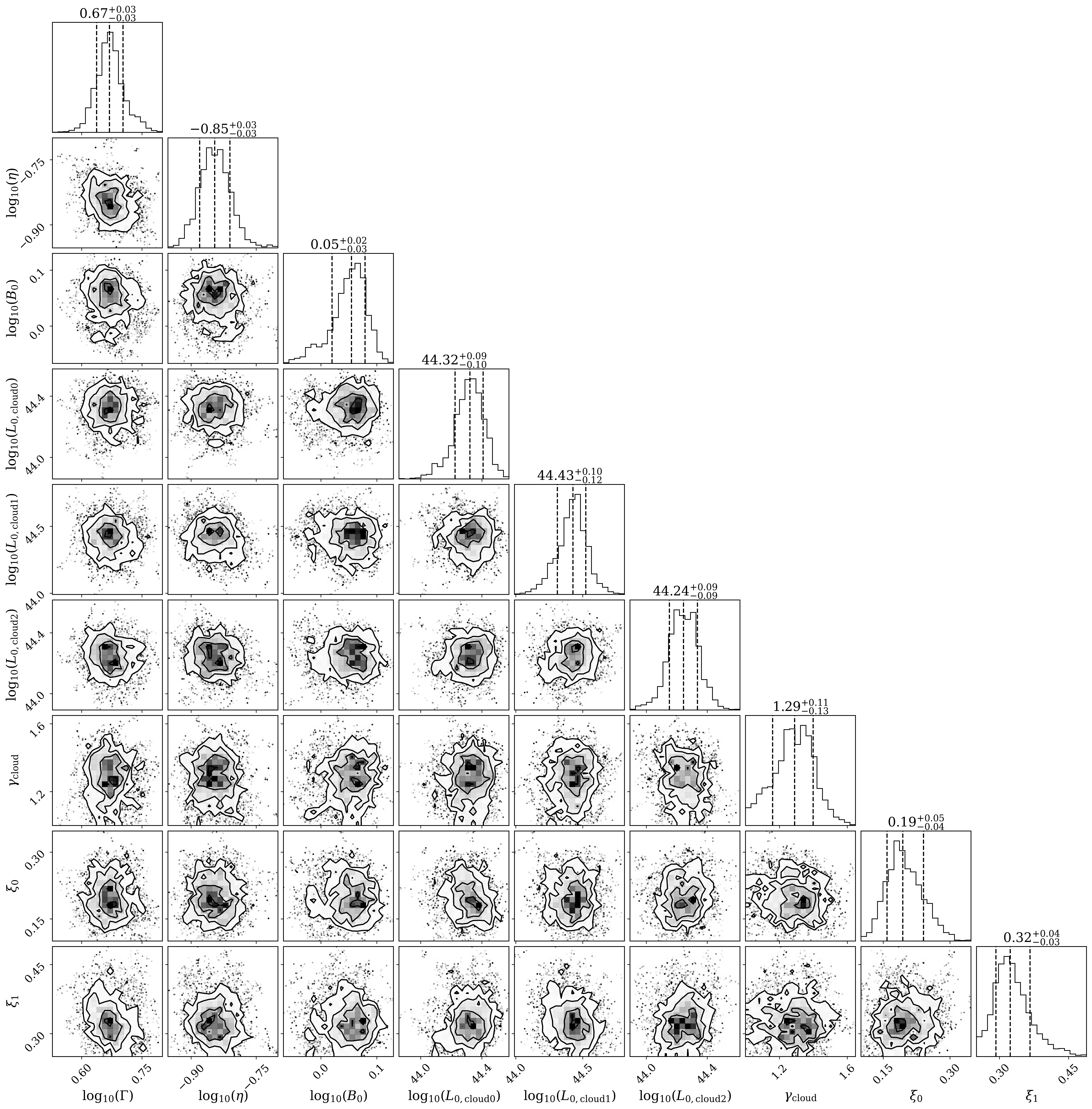  }
    \caption{ Parameter values relevant to the radio clouds explored through MCMC sampling. We plot the distributions of parameter values sampled through our MCMC study, reflecting $\sim$40,000 realizations. This represents exploration of the parameter space near a local, high-likelihood solution. Contours represent the 1, 2, and 3$\sigma$ credible regions in two dimensions.  }
    \label{fig:res_triangle_cloud}
\end{figure*}

In Fig. \ref{fig:res_triangle_cloud}, we provide the posterior distributions of the parameters describing the parsec-scale jet (cf. Sec. \ref{sec:model_radio_jet}) delayed radio flare. For visualization purposes, we repeat here the jet parameters $\Gamma, \beta$ and $B$. 

\section{Parameterization of the steady-state emission}
\label{app:steady}

In the main text, we described how we fit the 2017 multi-wavelength flare of blazar TXS~0506+056. The model describes the time-dependent evolution of an enhancement of the particle injection from the inner jet out to the parsec-scale; meanwhile, the rest of the jet also radiates, emitting a baseline multi-wavelength flux that we then add to the time-dependent flux from the flare before fitting to the data. For simplicity, we assume this baseline flux to be constant [cf. Eq.~(\ref{eq:fluxes})], as it results from steady-state emission from the extended jet. With this assumption, the entire temporal variability during the 2017 flare is described by our sequence of moving blobs.

\begin{figure}[ht]
    \centering
    \includegraphics[width=\linewidth]{ 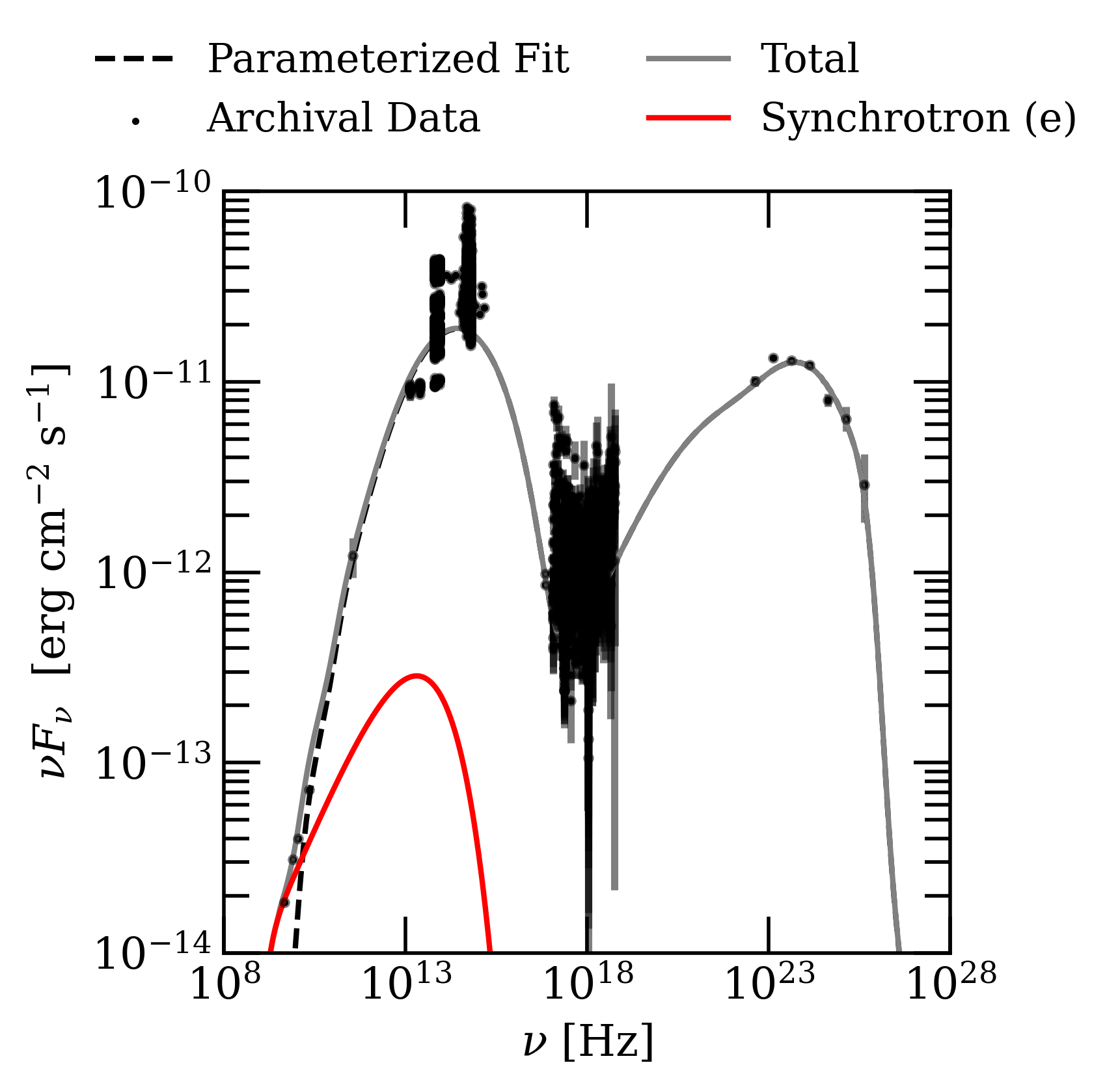 }
    \caption{Steady-state spectral model underlying our analysis of the evolving emission from TXS 0506+056. The gray curve shows the steady-state spectrum from the extended jet obtained in a previous work \citep{Rodrigues:2025cpm}, fitted to archival source data, shown as black data points. Since the model undershoots the radio emission, we additionally considered synchrotron emission from a second population of steady-state electrons, shown in red. The sum of the two, shown as a solid gray curve, corresponds to the baseline steady-state spectrum used in this work.}
    \label{fig:steady}
\end{figure}

We adopt a recent result from a leptohadronic extended jet model applied to TXS~0506+056 \citep{Rodrigues:2025cpm} for the spectral shape of the steady-state baseline emission. This is shown as a gray curve in Fig. \ref{fig:steady}, together with the archival multi-wavelength data, shown here as black data points. In the gamma-ray range, the data represent the average Fermi-LAT spectrum, about a factor of five lower than the peak gamma-ray flux during the 2017 flare. Although this extended jet model predicts higher levels of radio emission than most single-zone models, owing to the inclusion of a continuous particle population out to the parsec-scale jet, we can see by the gray curve that it still undershoots the radio fluxes below some tens of GHz. To better describe the baseline radio flux, we also consider steady-state synchrotron emission from an additional electron population, shown as a red curve in Fig. \ref{fig:steady}. We perform a by-eye fit of the archival data, assuming contributions from both the existing model description and a one-zone model of electrons with variable parameters. The additional, contributing electron population has a bulk Lorentz factor of 10, magnetic field strength of 0.03 G, and a radius of 2000 light days. The electron spectrum has an index $\gamma = 2.1$, a normalization, $1.9 \times 10^{41}$ erg/s, and minimum and maximum electron energies of 1.0 and $6.0 \times 10^{3}$ in units of energy relative to the electron rest mass.

The sum of the two spectra is shown as a black curve. We adopt this total spectrum as the constant steady-state flux $dN_\mathrm{steady}/(dE_\gamma dt)$ in Eq.~(\ref{eq:fluxes}).

\bibliographystyle{apsrev4-2}
\bibliography{neutrino}
 
\end{document}